\documentclass[graybox]{svmult}

\usepackage{type1cm}        
\usepackage{makeidx}         
\usepackage{graphicx}        
\usepackage{multicol}        
\usepackage[bottom]{footmisc}

\usepackage{newtxtext}       %
\usepackage[varvw]{newtxmath}       

\makeindex             

\usepackage{bm}
\usepackage{caption}
\usepackage{cite}

\def\be {\begin{eqnarray}}
\def\ee {\end{eqnarray}}
\def\E {{\bf E}}
\def\H {{\bf H}}
\def\r {{\bf r}}
\def\R {{\bf R}}
\def\k {{\bf k}}
\def\kk {{\bm \kappa}}
\def\pp {_\parallel}
\def\ki {{\bf k}_\text{i}}
\def\khati {\hat{\bf k}_\text{i}}
\def\kipp {{\bf k}_{\text{i},\parallel}}
\def\khats {\hat{\bf k}_\text{s}}
\def\kspp {{\bf k}_{\text{s},\parallel}}
\def\ks {{\bf k}_\text{s}}
\def\ksz {k_{\text{s},z}}
\def\kiz {k_{\text{i},z}}

\def\p {{\bf p}}
\def\f {{\bf f}}

\def\rj {{\bf r}_j}
\def\rl {{\bf r}_l}
\def\rjpp {{\bf r}_{j,\parallel}}

\def\Eb {{\bf E}_\text{b}}
\def\Ee {{\bf E}_\text{exc}}
\def\ei {\hat{\bm e}_\text{i}}

\def\er {\hat{\bm e}_\text{r}}
\def\et {\hat{\bm e}_\text{t}}
\def\Es {{\bf E}_\text{s}}
\def\Gb {{\bf G}_\text{b}}
\def\G {{\bf G}}
\def\P {{\bf P}}
\def\gb {{\bf g}_\text{b}}
\def\T {{\bf T}}

\def\Tl {{\bf T}_l}
\def\Tj {{\bf T}_j}

\def\kb {k_\text{b}}
\def\u {{\bf u}}
\def\Ig {{\bf I}}

\def\epsb {{\epsilon_\text{b}}}
\def\ll {\left\langle}
\def\rr {\right\rangle}
\def\deps {\delta{\bm \epsilon}}

\begin{document}

\title*{Specular reflection and transmission of electromagnetic waves by disordered metasurfaces}

\author{Kevin Vynck, Armel Pitelet, Louis Bellando and Philippe Lalanne}

\institute{Kevin Vynck \at Universit\'{e} Claude Bernard Lyon 1, CNRS, iLM, F-69622 Villeurbanne, France, \email{kevin.vynck@univ-lyon1.fr}
\and Armel Pitelet \at Universit\'{e} de Bordeaux, Institut d'Optique Graduate School, CNRS, LP2N, F-33400 Talence, France, \email{armel.pitelet@gmail.com}
\and Louis Bellando \at Universit\'{e} de Bordeaux, CNRS, LOMA, F-33400 Talence, France, \email{louis.bellando.physique@gmail.com}
\and Philippe Lalanne \at Universit\'{e} de Bordeaux, Institut d'Optique Graduate School, CNRS, LP2N, F-33400 Talence, France, \email{philippe.lalanne@institutoptique.fr}}
%
%
\maketitle

\abstract{
Planar, disordered assemblies of small particles incorporated in layered media -- sometimes called ``disordered metasurfaces'' in the recent literature -- are becoming widespread in optics and photonics. Their ability to scatter light with exotic angular and spectral features in reflection and transmission, as well as their suitability to scalable fabrication techniques, makes them promising candidates for certain applications, ranging from thin-film photovoltaics to visual appearance design. This chapter introduces the basic concepts and theoretical models for the specular (a.k.a. coherent) reflectance and transmittance of electromagnetic waves by disordered metasurfaces. After describing the classical scattering formalism for discrete media, we establish known analytical expressions for the reflection and transmission coefficients of disordered particle monolayers on layered substrates. Two classical models, based on the independent scattering approximation (ISA) and the effective field approximation (EFA), are presented. Their accuracy is examined by comparing predictions with those obtained from rigorous full-wave computations using an in-house multiple-scattering code. This chapter may serve as a starting point to students and researchers who wish to dive into the topic and explore the potential of disordered metasurfaces for applications.
}


\section{Introduction}
\label{sec:intro}

The phenomenon of wave scattering by random rough surfaces is encountered in many physical settings and techniques. These include radar remote sensing and imaging of planetary surfaces, optical characterization of surfaces and LIDAR, underwater acoustics, radio communications, and seismology, to cite only a few. A typical problem of interest is a coherent wave impinging on a surface with random heterogeneities (e.g., random height variations, supported particles at random positions, etc.). The intensity of the reflected signal generally takes the form of a speckle, showing bright and dark spots due to constructive and destructive interferences between waves coming from different positions of the sample. As might be expected, the statistical properties of the measured signal contain statistical information on the surface topology. Considerable advances have been made in the past decades on establishing this relationship, as discussed in several excellent textbooks~\cite{beckmann1987scattering, stover1995optical, bedeaux2004optical, vesperinas2006scattering, voronovich2013wave}.

This period has also been marked by the exploration of novel strategies to control the emission, propagation and confinement of light with nanostructures -- a discipline nowadays known as ``nanophotonics''~\cite{benisty2022introduction}. Research has been greatly stimulated by advances in nanofabrication techniques, which enabled the realization of a large variety of dielectric and metallic subwavelength structures with nanometer-scale resolution, in one, two and three dimensions~\cite{lopez2003materials, mascolo2011production}. One major outcome has been to unveil the potential of finely-engineered high-index resonant nano-objects to exhibit exotic properties, such as a light bending effect~\cite{mirin2009light}, a strong scattering anisotropy~\cite{gomez2011electric} or a side-dependent coupling to waveguides~\cite{wu2020intrinsic}. By optimizing the composition, size and shape of the nano-objects and placing them on a substrate, one creates a so-called ``metasurface'' that can deviate, focus, or alter the polarization state of an incident beam. The topic has become extremely popular and is covered by many review and expert opinion articles~\cite{kuznetsov2016optically, lalanne2017metalenses, kamali2018review, shaltout2019spatiotemporal}.

While in most situations, the nano-objects are ordered, i.e., arranged deterministically on a lattice, and the surfaces are realized by top-down fabrication techniques (e.g., lithography), increasing efforts are being made to realize metasurfaces by \textit{bottom-up} techniques, like colloidal chemistry and self-assembly. A major motivation is the expected benefit in terms of scalability and manufacturing cost. The resulting structures are usually disordered, even though some degree of control over structural correlations is possible~\cite{roach2022controlling}. Light interaction with disordered metasurfaces is receiving growing attention since a few years, with a focus on functionalities that do not require a precise control over the position of the inclusions, such as light absorption for photodetection, sensing and photovoltaics~\cite{moreau2012controlled, chevalier2015absorbing, stewart2017toward, piechulla2021antireflective}, light extraction for organic LEDs~\cite{jouanin2016designer, donie2021planarized}, and light scattering for augmented reality displays~\cite{bertin2018correlated} and visual appearance design~\cite{vynck2022visual}.

Disordered metasurfaces may thus be seen as a special type of random scattering surface and, naturally, one is led to the question of whether (and how well) previously-established theoretical models can be used for the purpose. The present chapter aims at answering, at least in part, to this question.

In modelling studies, the intensity scattered by a surface is generally decomposed into two components: a so-called \textit{coherent} component, which corresponds to the specularly reflected and transmitted waves\footnote{The term ``specular'' is used here independently for both reflected and transmitted waves, but one may also use the term ``ballistic'' for the transmitted waves.}, and a so-called \textit{incoherent} component, which corresponds to the diffuse (non-specular) intensity. Formally, as will be shown below, the former is given by the ensemble-averaged electric field produced by the system and leads to the definition of complex reflection and transmission coefficients, whereas the latter is given by the fluctuations of the electric field around its average value and leads to the definition of an angle-resolved scattering diagram. The theoretical modelling of the diffuse intensity created by planar, disordered assemblies of resonant particles in layered media remains, to our knowledge, an open problem, as models are in general limited to particle monolayers embedded in a uniform background~\cite{loiko2018incoherent} or to particles much smaller than the wavelength~\cite{sasihithlu2016surface}. We will therefore focus here on the specular response of disordered metasurfaces, for which a more exhaustive literature exists.

The problem of determining the reflection and transmission coefficients of particle monolayers is closely related to that of the electromagnetic homogenization. For particles that are very small compared to the wavelength in the embedding medium (i.e., in the quasi-static regime), the system may be seen as a homogeneous layer with an effective permittivity tensor\footnote{The effective permittivity of a monolayer of particles is not a scalar in general.}, which may be obtained from the Maxwell-Garnett mixing rule~\cite{sasihithlu2016surface}, the Yamaguchi model~\cite{yamaguchi1974optical} or its extensions~\cite{fedotov2003optical, toudert2012advanced}. An alternative approach, still applicable to monolayers of very small particles only, proposed by Bedeaux and Vlieger~\cite{bedeaux1973phenomenological, vlieger1980statistical} and implemented in the software GranFilm~\cite{lazzari2002granfilm}, relies on the notions of excess fields and surface susceptibilities to solve the electromagnetic problem. However, for larger particles possibly exhibiting high-order resonances, one needs to resort to the electromagnetic scattering theory by discrete media~\cite{tsang2004scattering}, an approach that has notably been followed by Garc\'{i}a-Valenzuela and colleagues~\cite{pena2006coherent, garcia2012multiple}.

\begin{figure}[h!]
	\centering
	\includegraphics[scale=1]{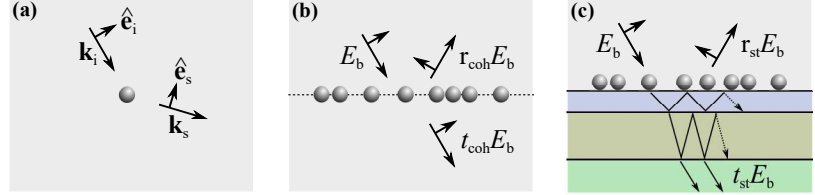}
	\caption{Illustration of the different problems of interest in this chapter and definition of certain variables. A discrete medium composed of particles is illuminated by a planewave $\Eb$ with amplitude $E_\text{b}$, wavevector $\ki$ and polarization $\ei$. (a) Electromagnetic scattering by an isolated particle in a uniform medium. In Section~\ref{sec:basic-formalism}, we will derive an expression of the scattered field in a planewave basis. (b) Specular reflection and transmission by a planar, disordered assembly of identical particles in a uniform medium. Section~\ref{sec:monolayers-theory} aims mainly at presenting two models for the complex reflection and transmission coefficients of particle monolayers in uniform media, $r_\text{coh}$ and $t_\text{coh}$, respectively. (c) Specular reflection and transmission by a disordered metasurface made of identical particles on a layered substrate. We will explain, at the end of Section~\ref{sec:monolayers-theory}, how the models can be used to predict the specular reflection, described by the coefficient $r_\text{st}$, of particle monolayers on layered substrates. In Section~\ref{sec:numerical-aspects}, we will test those models using full-wave multiple-scattering computations.}
	\label{fig:definition}
\end{figure}

In light of the increasing attention given to disordered metasurfaces in recent years, we propose in this chapter to give some keys to understand the physical origin and underlying approximations, and test the validity of known analytical expressions for the reflection and transmission coefficients of particle monolayers in layered media. The chapter is decomposed as follows. In Section~\ref{sec:basic-formalism}, starting from the simple case of an individual, finite-size heterogeneity in a uniform medium [Fig.~\ref{fig:definition}(a)], we introduce the basic theoretical concepts in electromagnetic scattering, namely the wave propagation equation and the transition operator, the dyadic Green function and its angular spectrum representation, and the scattering amplitude. Eventually, we arrive at an expression of the field scattered by a particle in a planewave basis. In Section~\ref{sec:monolayers-theory}, we treat the problem of the coherent intensity scattered by a monolayer of particles in a layered geometry. After introducing the multiple-scattering equations and formally defining the coherent and incoherent intensities, we present two models for the specular reflection and transmission by (infinite) monolayers of particles in a uniform medium [Fig.~\ref{fig:definition}(b)]. We conclude this section with a brief explanation of how the results can be generalized to handle particle monolayers above a layered substrate [Fig.~\ref{fig:definition}(c)]. In Section~\ref{sec:numerical-aspects}, finally, we test the validity of these two analytical models on different systems, made of either metallic or dielectric particles, on a bare semi-infinite or layered substrate, by comparing the analytical predictions with those from rigorous, full-wave computations using an in-house multiple-scattering code.

\section{Basics of electromagnetic scattering by particles}
\label{sec:basic-formalism}

\subsection{Wave equations}

We consider a finite region of space filled with a non-magnetic material (relative permeability, $\mu(\mathbf{r}) = 1$), described by a relative permittivity ${\bm \epsilon}(\r)$ in a uniform \textit{host} medium with relative scalar permittivity $\epsb$. The permittivity variation $\deps(\r) \equiv ({\bm \epsilon}(\r)-\epsb\Ig)$, with $\Ig$ the unit tensor, defines a compact heterogeneity, which will be our particle lateron, though no assumption on the particle composition, size or shape is in fact necessary. We consider harmonic fields at frequency $\omega$ with the $e^{-i \omega t}$ convention and drop hereafter the explicit dependence in the permittivities, fields, etc. for simplicity.

We start from the macroscopic Maxwell's (curl) equations for the electric and magnetic fields, $\E$ and $\H$, at frequency $\omega$ with a current density source $\mathbf{J}$,
\be
  \bm{\nabla} \times \E(\r) &=& i \omega \mu_0 \H(\r), \label{eq:maxwell-1} \\
  \bm{\nabla} \times \H(\r) &=& -i \omega \epsilon_0 \bm{\epsilon}(\r) \E(\r) + \mathbf{J}(\r).\label{eq:maxwell-2}
\ee
Taking the curl of Eq.~\eqref{eq:maxwell-1} and inserting Eq.~\eqref{eq:maxwell-2} in the resulting expression leads to a vector wave propagation equation for the electric field
\be
  \bm{\nabla} \times \bm{\nabla} \times \E(\r) - \frac{\omega^2}{c^2} \bm{\epsilon}(\r) \E(\r) =  i \omega \mu_0 \mathbf{J}(\r), \label{eq:wave-equation-total}
\ee
where $c^2=(\epsilon_0 \mu_0)^{-1}$, $c$ being the speed of electromagnetic waves in vacuum.

In a scattering problem, it is convenient to decompose the field as the sum of a background field and a scattered field, $\E = \Eb + \Es$. The background field $\Eb$ is the solution of the wave propagation equation with the source term but without the heterogeneity,
\be
  \bm{\nabla} \times \bm{\nabla} \times \Eb(\r) - \kb^2 \Eb(\r) =  i \omega \mu_0 \mathbf{J}(\r), \label{eq:wave-equation-background}
\ee
with $\kb^2=k_0^2 \epsb$ and $k_0=\omega/c$. From Eqs.~\eqref{eq:wave-equation-total} and \eqref{eq:wave-equation-background}, one therefore reaches a wave equation for the scattered field only,
\be
  \bm{\nabla} \times \bm{\nabla} \times \Es(\r) - \kb^2 \Es(\r) = k_0^2 \deps(\r) \E(\r). \label{eq:wave-equation-scattered}
\ee
The scattered field in the background medium is thus generated by the total field in the heterogeneity volume, which is nothing but the polarization density $\P(\r)/\epsilon_0 = \deps(\r) \E(\r)$.

\subsection{Lippmann-Schwinger equation}

The electromagnetic problem described by Eq.~\eqref{eq:wave-equation-scattered} can conveniently be reformulated by introducing the notion of dyadic Green function~\cite{tai1994dyadic}. The dyadic Green function $\G(\r,\r')$ of a medium describes the electric field produced at point $\r$ by a radiating point electric dipole at point $\r'$. While it can be defined for an arbitrary environment, we are interested here in the dyadic Green function in the background medium, $\Gb$, which is then the solution of
\be
  \bm{\nabla} \times \bm{\nabla} \times \Gb(\r,\r') - \kb^2 \Gb(\r,\r') = \Ig \delta(\r-\r'). \label{eq:wave-equation-Green}
\ee
Multiplying both sides by $k_0^2 \deps(\r') \E(\r')$, integrating over $\r'$ and using Eq.~\eqref{eq:wave-equation-scattered} leads to
\be
  \Es(\r) = k_0^2 \int \Gb(\r,\r') \deps(\r') \E(\r')  d\r'. \label{eq:Es-Green-integral}
\ee
Only the field in the heterogeneity volume contributes to the scattered field. Thus, ones reaches the so-called Lippmann-Schwinger equation for the total field,
\be
	\E(\r) = \Eb(\r) + k_0^2 \int \Gb(\r,\r') \deps(\r') \E(\r')  d\r'. \label{eq:LS-integral}
\ee
The equation is physically very insightful, but not very practical as such because the field at $\r$ depends explicitly on the field at $\r'$.

\subsection{Transition operator}

To reach more practical expressions, we can express the scattered field in Eq.~\eqref{eq:Es-Green-integral} by successive iterations as a series of scattering events,
\be
  \Es(\r) &=& k_0^2 \int \Gb(\r,\r') \deps(\r') \Eb(\r')  d\r' \nonumber \\
  &+& k_0^4 \int \Gb(\r,\r') \deps(\r') \Gb(\r',\r'') \deps(\r'') \Eb(\r'') d\r' d\r'' \nonumber \\
  &+& ...
\ee
The first integral term gives the scattered field due to one interaction within the heterogeneity, the second to two interactions within the heterogeneity, etc. The scattered field is now expressed in terms of the background field only. Eventually, all multiple-scattering orders can be incorporated into a transition operator $\T$, hereafter called T-operator\footnote{$\T$ is called here the T-operator to avoid the confusion with the T-matrix, which, in the numerical method of the same name~\cite{mishchenko1996t}, is generally defined in a basis of vector spherical wave functions.}, such that
\be
  \Es(\r) = \int \Gb(\r,\r') \T(\r',\r'') \Eb(\r'') d\r' d\r''. \label{eq:Es_T-operator-1}
\ee
with
\be
  \T(\r,\r') = k_0^2 \deps(\r) \left[ \delta(\r-\r') + \int \Gb(\r,\r'') \T(\r'',\r') d\r'' \right], \label{eq:T-operator}
\ee
Equation~\eqref{eq:Es_T-operator-1} can be physically interpreted as follows: a background field $\Eb$ arriving a point $\r''$ in the heterogeneity induces, via the T-operator $\T$, a dipole moment at position $\r'$, which then radiates in the background medium towards point $\r$ via the dyadic Green function $\Gb$. Note that $\T$ is spatially non-local in general, as it contains all multiple-scattering events relating two points within the heterogeneity. It is a complicated, yet intrinsic quantity: once known, it allows predicting the scattered field for any background field.

Because the microscopic details of the heterogeneity are ``hidden'' in the T-operator, we will hereafter talk about particles explicitly when referring to an individual, finite-size heterogeneity.

The T-operator can be obtained analytically using Mie theory for simple objects (e.g., particles with spherical symmetry) and numerically by solving Maxwell's equations otherwise~\cite{bohren2008absorption}. Interestingly, for assemblies of particles, one can define a T-operator of the whole system, which can be expressed in terms of the T-operators of the individual particles. The computation can be achieved by solving the multiple-scattering problem using, for instance, so-called the T-matrix method~\cite{mishchenko1996t}.

\subsection{Dyadic Green function in a homogeneous medium}

To proceed further, we need to define more properly the dyadic Green function above. We will skip the derivation, which can be found elsewhere~\cite{tai1994dyadic, jackson1999classical, novotny2012principles, carminati2021principles}. In translationally-invariant and isotropic media, the dyadic Green function $\Gb(\r,\r') \equiv \Gb(\r-\r')$ in Cartesian coordinates is given by
\be
  \Gb(\r-\r') = - \frac{\delta(\R)}{3\kb^2} \Ig + \text{PV} \left\lbrace \frac{\exp[i \kb R]}{4\pi R} \left[ \Ig - \u \otimes \u + \frac{i \kb R-1}{(\kb R)^2} \left( \Ig - 3 \u \otimes \u \right) \right] \right\rbrace, \nonumber \\
  \label{eq:Green_hom_real}
\ee
where $\delta(.)$ is the Dirac delta function, $\text{PV}$ stands for Principal Value, $\otimes$ denotes the tensor product, and we defined $\R=\r-\r'=R\u$ for conciseness. The first term corresponds to the singularity of the dyadic Green function at the origin and the second term to the non-local part, which itself contains evanescent and propagating components. In the far-field, $\kb R \gg 1$, only propagating waves remain and the expression reduces to
\be
  \Gb(\r-\r') \sim \frac{\exp[i \kb R]}{4\pi R} \left[ \Ig - \u \otimes \u \right]. \label{eq:Green_hom_real_far}
\ee
Equation~\eqref{eq:Green_hom_real_far} shows that the field radiated by a compact heterogeneity behaves at large distances as a transverse, outgoing spherical wave, where the transverse nature is due to the term between brackets.

The representation in outgoing spherical waves is well suited to scattering problems in volumes but when the problem involves planar geometries (e.g., particle monolayers, interfaces), it is more convenient to express the waves as a linear combination of planewaves. This is the so-called angular spectrum representation, a.k.a. Weyl expansion or Weyl identity~\cite{weyl1919ausbreitung}, and that we will now apply to the dyadic Green function.

For this, let us start by expressing the dyadic Green function in real space in terms of its Fourier transform,
\be
  \Gb(\r-\r') = \frac{1}{(2\pi)^3} \int \frac{1}{k^2 - \kb^2} \left[ \Ig - \frac{\k \otimes \k }{\kb^2} \right] \exp \left[i \k \cdot (\r-\r') \right] d\k. \label{eq:Green_hom}
\ee
Defining $\r = \left[\r\pp, z \right]$, $\k = \left[ \k\pp, k_z \right]$, $k\pp=|\k\pp|$, and $\gamma = \sqrt{\kb^2 - k\pp^2}$ with $\text{Re}[\gamma] >0$ and $\text{Im}[\gamma] >0$, Eq.~\eqref{eq:Green_hom} can then be rewritten as
\be
  \Gb(\r-\r') = \frac{1}{(2\pi)^2} \int \gb(\k\pp,z,z') \exp \left[ i \k\pp \cdot (\r\pp-\r'\pp) \right] d\k\pp,\label{eq:Green_Fourier}
\ee
with
\be
  \gb(\k\pp,z,z') = \frac{1}{2\pi} \int \frac{1}{k_z^2 - \gamma^2} \left[ \Ig - \frac{\k \otimes \k }{\kb^2} \right] \exp \left[ i k_z (z-z') \right] dk_z.  \label{eq:Green_2DWeyl_integral}
\ee
The solution of the integral over $k_z$ in Eq.~\eqref{eq:Green_2DWeyl_integral} is evaluated taking care of the poles by using Cauchy's residue theorem to the contour integrals, leading to
\be
  \gb(\k\pp,z,z') = - \frac{\delta(z-z')}{\kb^2} \bm{\delta}_{zz} + \text{PV}_{zz} \left\lbrace \frac{i}{2} \frac{1}{\gamma} \left[ \Ig - \frac{\kk^\pm \otimes \kk^\pm }{\kb^2} \right] \exp\left[ i \gamma |z - z'| \right] \right\rbrace
\ee
where $\bm{\delta}_{ij}$ is the Kronecker delta tensor and $\kk^\pm = [\k\pp,\pm \gamma]$ with $\pm = \text{sign}(z - z')$. Inserting this expression into Eq.~\eqref{eq:Green_Fourier} and keeping the non-singular part only (i.e., $z \neq z'$), we obtain the following angular spectrum represenation of the dyadic Green function,
\be
  \Gb(\r-\r') = \frac{i}{2(2\pi)^2} \int \frac{1}{\gamma} \left[ \Ig - \frac{\kk^\pm \otimes \kk^\pm}{\kb^2} \right] \exp \left[ i \kk^\pm \cdot (\r-\r') \right] d\k\pp. \label{eq:Green_Weyl}
\ee
Note that the integral is made over the entire (infinite) reciprocal plane described by $\k\pp$. Because $\gamma = \sqrt{\kb^2 - k\pp^2}$, wavevectors fulfilling $k\pp^2 \leq \kb^2$ or $k\pp^2 > \kb^2$ correspond to propagating or evanescent waves, respectively.

\subsection{Scattering of a planewave by a particle}

To complete this section, we will now express the scattered field $\Es$ produced by an individual particle in the angular spectrum representation. Let us first consider that the particle is centered at position $\rj$ and is described by a T-operator $\Tj$. The relative position of an individual particle in space does not affect its scattering properties, of course, but the relative position between particles in an assembly does. This step is therefore important for our purpose, as will be shown lateron. Equation~\eqref{eq:Es_T-operator-1} then reads
\be
  \Es(\r) = \int \Gb(\r,\r') \Tj(\r'-\rj,\r''-\rj) \Eb(\r'') d\r' d\r''. \label{eq:Es_T-operator}
\ee
The T-operator can be expressed in terms of its Fourier transform as
\be
  \Tj(\r'-\rj,\r''-\rj) &=& \frac{1}{(2\pi)^6} \int \exp \left[ i \p' \cdot (\r' - \rj) \right] \nonumber \\
  &\times& \Tj (\p', \p'') \exp \left[ - i \p'' \cdot (\r'' - \rj) \right] d\p' d\p''. \label{eq:T-operator-Fourier}
\ee

Let us then consider that the background field $\Eb$ is a planewave with amplitude $E_\text{b}$, wavevector $\ki=\kb \khati$ and polarization $\ei$,
\be
  \Eb(\r) = E_\text{b} \ei \exp \left[ i \ki \cdot \r \right]. \label{eq:incident_planewave}
\ee
Using Eqs.~\eqref{eq:T-operator-Fourier} and \eqref{eq:Green_Weyl}, we can rewrite Eq.~\eqref{eq:Es_T-operator} as
\be
  \Es(\r) = \frac{i E_\text{b}}{2(2\pi)^8} &\int& \frac{1}{\gamma} \left[ \Ig - \frac{\kk^\pm \otimes \kk^\pm }{\kb^2} \right] \Tj (\p', \p'') \ei \exp \left[ i \kk^\pm \cdot (\r-\r') \right] \nonumber \\
  &\times& \exp \left[ i \p' \cdot (\r' - \rj) \right] \exp \left[ - i \p'' \cdot (\r'' - \rj) \right]  \exp \left[ i \ki \cdot \r'' \right] \nonumber \\
  &\times& d\r' d\r'' d\p' d\p'' d\k\pp.
\ee
Having that $\int \exp [i \mathbf{q} \r] d\r = (2\pi)^3 \delta(\mathbf{q})$, the integrals over $\r'$ and $\r''$, and then $\p'$ and $\p''$, lead to $\p' = \kk^\pm$ and $\p'' = \ki$. After simplification, we thus obtain
\be
  \Es(\r) = \frac{i E_\text{b}}{2(2\pi)^2} \exp \left[ i \ki \cdot \rj \right] &\int& \frac{1}{\gamma} \left[ \Ig - \frac{\kk^\pm \otimes \kk^\pm}{\kb^2} \right] \Tj (\kk^\pm,\ki) \ei \nonumber \\ 
  &\times& \exp \left[ i \kk^\pm \cdot (\r-\rj) \right] d\k\pp. \label{eq:Es_T_final}
\ee
Similarly to Eq.~\eqref{eq:Green_Fourier}, we can define $\Es(\k\pp,z,z_j)$ such that
\be
  \Es(\r) = \frac{1}{(2\pi)^2} \int \Es(\k\pp,z,z_j) \exp \left[ i \k\pp \cdot (\r\pp-\rjpp) \right]  d\k\pp,
\ee
leading to
\be
  \Es(\k\pp,z,z_j) &=& \frac{i E_\text{b}}{2} \frac{\exp[i \gamma |z-z_j|]}{\gamma} \left[ \Ig - \frac{\kk^\pm \otimes \kk^\pm}{\kb^2} \right] \Tj (\kk^\pm,\ki) \ei \exp[i \ki \cdot \rj]. \nonumber \\ \label{eq:Es-parallel-T}
\ee
Note that the wavevector $\k\pp$ still covers the entire reciprocal space, so that both evanescent and propagating waves are considered. As we are more interested here into waves propagating in the far field, let us select a propagating wave with wavevector $\ks=\kb \khats$ and consider a plane above the particle, $z>z_j$. We therefore use $\kk^+ = [\kspp,+\ksz]$. For waves propagating in the far field, we can further introduce the \textit{vector scattering amplitude}, a classical quantity in scattering theory~\cite{carminati2021principles}\footnote{Important quantities, notably the extinction cross-section and differential scattering cross-section, can be calculated directly from the scattering amplitude.},
\be
  \f_j(\khats,\khati) = \frac{1}{4\pi} \left[ \Ig - \khats \otimes \khats \right] \T_j (\kb \khats,\kb \khati). \label{eq:scattering-amplitude}
\ee
This leads to
\be
  \Es(\kspp,z,z_j) = 2i\pi E_\text{b} \frac{\exp[i \ksz (z-z_j)]}{\ksz} \f_j(\khats,\khati) \ei \exp[i \ki \cdot \rj], \label{eq:Es-parallel}
\ee
where $\ksz = \kb \cos(\theta_\text{i})$, with $\theta_\text{i}$ the angle of incidence with respect to the the $z$-axis. Equation~\eqref{eq:Es-parallel} formally expresses the planewave decomposition of the field scattered by a particle at $\r_j$ in the far field in terms of the particle scattering amplitude. Similar steps will later be made to derive expressions for the field scattered by a monolayer of particles and the resulting reflection and transmission coefficients.


\section{Specular reflection and transmission by particle monolayers}
\label{sec:monolayers-theory}

\subsection{Multiple scattering by discrete media}

In the previous section, we considered the electromagnetic scattering problem for an individual particle, centered at point $\rj$, described by a T-operator $\Tj$, and illuminated by a background field $\Eb$. In the present section, we will consider scattering by an \textit{ensemble} of such particles. In the general case, each particle may have a different composition, size and shape, and therefore a different T-operator. Following Eq.~\eqref{eq:Es_T-operator}, the field scattered by a set of $N$ particles is now simply the sum of the field scattered by each of them,
\be
  \Es(\r) = \sum_{j=1}^N \int \Gb(\r,\r') \Tj(\r'-\rj,\r''-\rj) \Ee^j(\r'') d\r' d\r''. \label{eq:Es_multiple_T-matrix}
\ee
We defined here $\Ee^j$ as the field exciting the $j$-th particle, which, very importantly, can differ from the background field $\Eb$. Indeed, quite intuitively, this exciting field for the $j$-th particle should be the sum of the background field and the field scattered by all other particles, $ l\neq j$, as
\be
  \Ee^j(\r) = \Eb(\r) + \sum_{l \neq j}^N \int \Gb(\r,\r') \Tl(\r'-\r_l,\r''-\r_l) \Ee^l(\r'') d\r' d\r''. \label{eq:Eexc_multiple_T-matrix}
\ee

Equations~\eqref{eq:Es_multiple_T-matrix} and \eqref{eq:Eexc_multiple_T-matrix} define the multiple scattering problem. The exciting field $\Ee^p$ depends on the positions and properties of all other particles and is generally unknown. This set of equations can be solved numerically on finite ensembles of particles using T-matrix methods~\cite{mishchenko1996t}, as mentioned above, however, the scattering response of a specific configuration of particles is generally of limited interest. Instead, one looks for statistical properties of the scattering response, such as the average field or intensity, intensity fluctuations, field-field or intensity-intensity correlations, etc~\cite{carminati2021principles}. Theoretically, the problem is tackled in general by taking the configurational average of the relevant quantity (the field, intensity, etc.) from Eqs.~\eqref{eq:Es_multiple_T-matrix} and \eqref{eq:Eexc_multiple_T-matrix}, and making certain approximations on the interaction between particles.

\subsection{Coherent and incoherent intensity}

The intensity scattered by a medium, which is the quantity that is usually measured experimentally, can formally be decomposed into two components, denoted as coherent and incoherent. To show this, let us write the scattered field as the sum of its average value and a fluctuating part,
\be
	\Es(\r) = \langle \Es(\r) \rangle  + \delta\Es(\r), \qquad \text{with } \ll \delta\Es(\r) \rr = 0.
\ee
The configurational average (formally defined below) is written here with angle brackets, $\langle \cdot \rangle$. Calculating the configurational average of the electric field norm squared, $\ll|\E(\r)|^2\rr$, which is directly proportional to the average intensity, shows that it can be decomposed into two terms,
\be
  \langle | \E |^2 \rangle &=& |\Eb|^2 + 2 \text{Re}[\Eb \cdot \langle \Es^* \rangle] + |\langle \Es \rangle|^2 + \langle | \delta\Es |^2 \rangle, \nonumber \\
  &=& |\langle \E \rangle|^2 + \langle | \delta\Es |^2 \rangle. \label{eq:coherent-incoherent}
\ee

The first term, $|\langle \E \rangle|^2 = |\Eb|^2 + 2 \text{Re}[\Eb \cdot \langle \Es^* \rangle] + |\langle \Es \rangle|^2$, is known as the \textit{coherent} intensity\footnote{$|\langle \E \rangle|^2$ contains an interference term between the background and scattered fields, hence the term ``coherent''.}. In volume scattering, this component defines the extinction coefficient of the medium, which describes the attenuation rate of an incident wave due to scattering and absorption. In surface scattering, this leads to specularly reflected and transmitted waves defined by their reflection and transmission coefficients.

The second term, $\langle | \delta\Es |^2 \rangle = \langle | \E |^2 \rangle - |\langle \E \rangle|^2 = \langle | \Es |^2 \rangle - |\langle \Es \rangle|^2$, is known as the \textit{incoherent} intensity\footnote{The term ``incoherent'' may be misleading since the diffuse intensity can be impacted by interference between scattered waves, as in the case of correlated disorder~\cite{vynck2021light}}. It corresponds to the diffuse intensity, created by field fluctuations from realization to realization and characterized by a differential scattering cross-section.

As stated previously, we are interested here in the specular reflection and transmission of a monolayer of particles, and will therefore focus on theory for the average field. For further details on the theory for the intensity, we recommend the textbooks from Tsang and Kong~\cite{tsang2004scattering} and from Carminati and Schotland~\cite{carminati2021principles}.

\subsection{Average scattered field}

We start by defining the configurational average for discrete media. Our main ingredient here is the $N$-dimensional probability density function $p(\R)$ of finding $N$ particles in the configuration $\R=[\r_1, \r_2,\cdots,\r_N]$. The configurational average of a variable $\bm{f}(\r)$ is defined as
\be
  \langle \bm{f}(\r) \rangle &=& \int \bm{f}(\r,\R) p(\R) d\R, \label{eq:configurational-average-def}
\ee
with $\bm{f}(\r,\R)$ the variable evaluated at $\r$ for a specific configuration $\R$.

Let us then proceed by calculating the configurational average of the scattered field, Eq.~\eqref{eq:Es_multiple_T-matrix}, considering a set of $N$ \textit{identical} particles described by a unique T-operator, $\Tj(\r'-\r_j,\r''-\r_j) \equiv \T(\r'-\r_j,\r''-\r_j)$. We thus have
\be
  \langle \Es(\r) \rangle = \sum_{j=1}^N \int \Gb(\r,\r') \left\langle \T(\r'-\rj,\r''-\rj) \Ee^j(\r'') \right\rangle d\r' d\r'', \label{eq:Es_multiple_T-matrix-identical}
\ee
with
\be
  \left\langle \T(\r'-\rj,\r''-\rj) \Ee^j(\r'') \right\rangle = \int \T(\r'-\rj,\r''-\rj) \Ee^j(\r'',\R) p(\R) d\R, \nonumber \\
\ee
following Eq.~\eqref{eq:configurational-average-def}. Using $p(\R)=p(\R|\rj)p(\rj)$ with $p(\R|\rj)$ the conditional probability density function for $\R$ having $\r_j$ fixed, and $d\R=d\r_1 d\r_2 \cdots d\r_N$, we obtain
\be
  \left\langle \T(\r'-\rj,\r''-\rj) \Ee^j(\r'') \right\rangle = \int \T(\r'-\rj,\r''-\rj) \langle \Ee^j \rangle_j (\r'',\r_j) p(\rj) d\rj, \nonumber \\ \label{eq:av-T-Exc-field-j}
\ee
where $\langle \Ee^j \rangle_j$ is the average exciting field $\Ee^j$ having the particle $j$ fixed at $\r_j$, given by
\be
  \langle \Ee^j \rangle_j (\r'',\r_j) = \int \Ee^j(\r'',\R) p(\R|\rj) d\r_1 \cdots d\r_{j-1} d\r_{j+1} \cdots d\r_N. \label{eq:Exc-field-j-fixed}
\ee
Inserting Eq.~\eqref{eq:av-T-Exc-field-j} into Eq.~\eqref{eq:Es_multiple_T-matrix-identical}, we finally obtain
\be
  \langle \Es(\r) \rangle = \sum_{j=1}^N \int \Gb(\r,\r') \T(\r'-\rj,\r''-\rj) \langle \Ee^j \rangle_j (\r'',\r_j)  p(\rj) d\rj d\r' d\r''. \label{eq:Es_multiple_Exc-field-j}
\ee

The difficulty at this stage is to express $\langle \Ee^j \rangle_j$ in a suitable manner as to solve Eq.~\eqref{eq:Es_multiple_Exc-field-j}. Without any approximation, $\langle \Ee^j \rangle_j$ can be expressed, in a similar manner as done above using Eq.~\eqref{eq:Eexc_multiple_T-matrix}, in terms of the average exciting field with two particles $j$ and $l$ fixed, and so on. Different levels of approximations are obtained depending on the order at which the truncation is made~\cite{tsang2004scattering}. Hence, the lowest-order approximation, known as independent scattering approximation (ISA), completely neglects the interaction between particles, assuming then $\langle \Ee^j \rangle_j (\r'',\rj) \simeq \Eb (\r'')$. Instead, the so-called effective field (or Foldy's) approximation (EFA) considers that $\langle \Ee^j \rangle_j (\r'',\rj) \simeq \langle \E \rangle (\r'')$, whereas the so-called quasi-crystalline approximation (QCA) assumes that $\langle \Ee^j \rangle_{jl} (\r'',\rj,\rl) \simeq \langle \Ee^j \rangle_j (\r'',\rj)$. We will treat below the ISA and EFA models. The QCA was used by Garc\'ia-Valenzuela \textit{et al.}~\cite{garcia2012multiple} and constitutes, to our knowledge, the state-of-the-art on multiple-scattering models for the coherent intensity from particle monolayers.

\subsection{Independent scattering approximation (ISA)}

The ISA assumes, as the name indicates, that the particles behave independently from each other; an electromagnetic wave interacts only once with each particle, the mutual interactions between particles are therefore neglected. The exciting field [Eq.~\eqref{eq:Eexc_multiple_T-matrix}] is then given by the background field only. Evidently, this approximation should only be valid for very dilute systems. Assuming then
\be
  \langle \Ee^j \rangle_j (\r'',\r_j) \simeq \Eb(\r''),
\ee
Eq.~\eqref{eq:Es_multiple_Exc-field-j} becomes
\be
  \langle \Es(\r) \rangle = \sum_{j=1}^N \int \Gb(\r,\r') \T(\r'-\rj,\r''-\rj) \Eb(\r'') p(\rj) d\rj d\r' d\r''.
\ee
Using Eqs.~\eqref{eq:Green_Weyl}, \eqref{eq:T-operator-Fourier} and \eqref{eq:incident_planewave}, and having $\int \exp[i(\p'-\kk^\pm) \cdot \r'] d\r' = (2\pi)^3 \delta(\p'-\kk^\pm)$ and $\int \exp[i(\ki-\p'') \cdot \r''] d\r'' = (2\pi)^3 \delta(\ki-\p'')$ leads to
\be
  \langle \Es(\r) \rangle = \frac{iE_\text{b}}{2(2\pi)^2} &\int& \frac{1}{\gamma} \left[ \Ig - \frac{\kk^\pm \otimes \kk^\pm}{\kb^2} \right] \T(\kk^\pm,\ki) \exp[i \kk^\pm \cdot \r] \ei \nonumber \\
  &\times& \left\lbrace \sum_{j=1}^N \int \exp[i(\ki-\kk^\pm) \cdot \r_j] p(\r_j) d\r_j \right\rbrace d\k\pp \label{eq:Es_ISA_curled}
\ee

Considering that the particles are randomly distributed on a surface of area $S$ at $z=0$ (for simplicity) leads to $p(\r_j) = \delta(z_j)/S$. In addition, since all particles obey the same statistics, the sum reduces to a prefactor $N$. Then, we consider the limit of an infinite surface, $\lim_{N,S \rightarrow \infty} (N/S) = \rho$, with $\rho$ the surface density. In this limit, the term between curled brackets in Eq.~\eqref{eq:Es_ISA_curled} becomes
\be
  \lim_{N,S \rightarrow \infty} \left\lbrace \frac{N}{S} \int \exp[i(\ki-\kk^\pm) \cdot \r_j] \delta(z_j) d\r_j \right\rbrace = (2\pi)^2 \rho \delta(\kipp-\k\pp),
\ee
thereby imposing the conservation of the parallel wavevector between the incident and scattered waves. Two solutions appear, $\kk^\pm = [\kipp,\pm \kiz] \equiv \k_\text{t/r}$, which correspond to the specularly reflected and transmitted planewaves. Having $\kiz = \kb \cos(\theta_\text{i})$ and using Eq.~\eqref{eq:scattering-amplitude} for the scattering amplitude, we reach
\be
  \lim_{N,S \rightarrow \infty} \langle \Es(\r) \rangle &=&  E_\text{b} \rho \frac{2i \pi}{\kb \cos(\theta_\text{i})} \f(\hat{\k}_\text{t/r},\khati) \ei \exp \left[ i \k_\text{t/r} \cdot \r \right], \label{eq:average-scattered-ISA} \\ 
  & \equiv & \langle \Es(\hat{\k}_\text{t/r}) \rangle \exp \left[ i \k_\text{t/r} \cdot \r \right].
\ee
Note the similarity with Eq.~\eqref{eq:Es-parallel} obtained for an individual particle, which makes sense considering the independent scattering approximation made here. Indeed, independent scattered waves emerging from random (but statistically uniform) positions on a surface add up coherently to form a planewave whose amplitude depends only on the scattering properties of the individual particle (i.e., the source of the scattered field) and the particle density.

Having now the average scattered field as a planewave with parallel wavevector $\kipp$ either in reflection or in transmission, we may define the complex reflection and transmission coefficients, $r_\text{coh}$ and $t_\text{coh}$, of the particle monolayer, as
\be
  r_\text{coh} E_\text{b} &=& \er \cdot \ll \Es(\hat{\k}_\text{r}) \rr, \label{eq:r_coh_def}\\
  t_\text{coh} E_\text{b} &=& E_\text{b} + \et \cdot \ll \Es(\hat{\k}_\text{t}) \rr, \label{eq:t_coh_def}
\ee
with $\er$ and $\et$ the polarization vectors of the specularly reflected and transmitted planewaves, respectively. Noting that $\k_\text{t}=\ki$, we obtain
\be
  r_\text{coh} &=& \rho  \frac{2 i \pi }{\kb \cos(\theta_\text{i})} \left( \er \cdot \f(\hat{\k}_\text{r},\khati) \ei \right), \label{eq:refl_coeff_ISA}\\
  t_\text{coh} &=& 1 + \rho \frac{2i \pi }{\kb \cos(\theta_\text{i})} \left( \et \cdot \f(\khati,\khati) \ei \right). \label{eq:trans_coeff_ISA}
\ee

For spherical particles, for which the scattering response depends on the scattering angle rather than the incident and scattered angles, and producing no polarization conversion ($\et = \ei$), these expressions can further be simplified using the following definitions
\be
  \f(\khati,\khati) \ei &\equiv& f(0) \ei, \label{eq:scatt_amplitude-1}\\
  \f(\hat{\k}_\text{r},\hat{\k}_\text{r}) \er &\equiv& f(0) \er, \label{eq:scatt_amplitude-2}\\
  \f(\hat{\k}_\text{r},\khati) \ei &\equiv& f_p(\pi-2\theta_\text{i}) \er, \label{eq:scatt_amplitude-3}\\
  \f(\khati,\hat{\k}_\text{r}) \er &\equiv& f_p(\pi-2\theta_\text{i}) \ei, \label{eq:scatt_amplitude-4}
\ee
where $f(0)$ is the forward scattering amplitude and $f_p(\pi-2\theta_\text{i})$ is the polarization-dependent scattering amplitude in the direction of specular reflection. Equations~\eqref{eq:refl_coeff_ISA} and \eqref{eq:trans_coeff_ISA} thus become
\be
  r_\text{coh} &=& \rho \frac{2 i \pi}{\kb \cos(\theta_\text{i})} f_p(\pi-2\theta_\text{i}), \label{eq:refl_coeff_ISA_spheres}\\
  t_\text{coh} &=& 1 + \rho \frac{2i \pi}{\kb \cos(\theta_\text{i})} f(0). \label{eq:trans_coeff_ISA_spheres}
\ee

We have therefore derived first equations for the specular reflection and transmission coefficients by particle monolayers, assuming here that the particles do not interact with each other. These expressions can be found in the literature on metasurfaces. Let us emphasize, however, that the coefficients diverge at grazing angles ($\theta_\text{i}$ approaching $\pi/2$). Intuitively, indeed, the interaction between particles should become more significant at oblique incidence, as the particles start to ``shadow'' each other. As we will see in Section~\ref{sec:numerical-aspects}, it can nevertheless be a good approximation for dilute media (1\% surface coverage).

\subsection{Effective field approximation (EFA)}

The EFA is a higher-order approximation than the ISA, in the sense that the average exciting field on particle $j$ (with particle $j$ fixed) is approximated not only by the background field but also the average scattered field, which partly accounts for the interaction between particles. Assuming thus
\be
  \langle \Ee^j \rangle_j (\r'',\r_j) \simeq \langle \E \rangle (\r''),
\ee
Eq.~\eqref{eq:Es_multiple_Exc-field-j} reads
\be
  \langle \Es(\r) \rangle = \sum_{j=1}^N \int \Gb(\r,\r') \T(\r'-\rj,\r''-\rj) \langle \E \rangle (\r'') p(\rj) d\rj d\r' d\r''.
\ee
Having $\langle \E \rangle = \Eb + \langle \Es \rangle$, this brings us to a similar problem as that with the Lippmann-Schwinger equation, Eq.~\eqref{eq:LS-integral}, namely that the field at a point $\r$ depends on the field at another point $\r''$. As previously, this can be solved by successive iterations, thereby expressing the problem as a series of scattering events between particles (not within a single heterogeneity as in the derivation of the T-operator).

A similar procedure as done above for the ISA can be followed and applied to the resulting infinite sum, leading to
\be
  \lim_{N,S \rightarrow \infty} \langle \Es(\r) \rangle &=& E_\text{b} \rho \frac{2i \pi}{\kb \cos(\theta_\text{i})} \f(\hat{\k}_\text{t/r},\khati) \ei \exp \left[ i \k_\text{t/r} \cdot \r \right]  \nonumber \\
  &+& E_\text{b} \rho^2 \left(\frac{2i \pi }{\kb \cos(\theta_\text{i})} \right)^2 \f(\hat{\k}_\text{t/r},\hat{\k}_\text{t/r}) \f(\hat{\k}_\text{t/r},\khati) \ei \exp \left[ i \k_\text{t/r} \cdot \r \right] \nonumber \\
  &+& \cdots
\ee
The first integral term of the sum involves one vector scattering amplitude and is identical to Eq.~\eqref{eq:average-scattered-ISA}, the second integral term involves two vector scattering amplitudes (note the different input and output wavevector directions), etc. The sum takes the form of a geometric series, which, using Eqs.~\eqref{eq:r_coh_def} and \eqref{eq:t_coh_def}, leads to new expressions for the reflection and transmission coefficients
\be
  r_\text{coh} &=& \er \cdot \left[ \Ig - \rho  \frac{2 i \pi }{\kb \cos(\theta_\text{i})} \f(\hat{\k}_\text{r},\hat{\k}_\text{r}) \right]^{-1} \left( \rho  \frac{2 i \pi }{\kb \cos(\theta_\text{i})} \right) \f(\hat{\k}_\text{r},\khati) \ei, \label{eq:refl_coeff_EFA}\\
  t_\text{coh} &=& \et \cdot \left[ \Ig - \rho  \frac{2 i \pi }{\kb \cos(\theta_\text{i})} \f(\khati,\khati) \right]^{-1} \ei, \label{eq:trans_coeff_EFA}
\ee
where $[\cdot]^{-1}$ denotes the dyadic inverse.

For spherical particles, we can use the equalities in Eqs.~\eqref{eq:scatt_amplitude-1}-\eqref{eq:scatt_amplitude-4} to reach
\be
  r_\text{coh} &=& \frac{ \rho  \frac{2 i \pi }{\kb \cos(\theta_\text{i})}  f_p(\pi-2\theta_\text{i})}{1-\rho  \frac{2 i \pi }{\kb \cos(\theta_\text{i})} f(0)}, \label{eq:refl_coeff_EFA_spheres}\\
  t_\text{coh} &=& \frac{1}{1-\rho  \frac{2 i \pi }{\kb \cos(\theta_\text{i})} f(0)}, \label{eq:trans_coeff_EFA_spheres}
\ee

We have thus reached a second set of equations for the reflection and transmission coefficients of particle monolayers, which considers the interaction between particles in a mean-field sense. Importantly, unlike in the ISA, the coefficients now behave correctly at grazing angles. Taking the limit $\theta_\text{i} \rightarrow \pi/2$ leads to a reflection coefficient $r_\text{coh}$ that approaches -1 (and thus, a reflectance $|r_\text{coh}|^2$ approaching 1) since $f_p(\pi-2 \theta_\text{i}) \rightarrow f(0)$, and a transmission coefficient $t_\text{coh}$ approaching 0. We will see in Section~\ref{sec:numerical-aspects} that the EFA leads to physically sound and quantitatively accurate predictions for moderately dense systems (typically, 10\% surface coverage) even at relatively large incident angles ($60^\circ$).

As a final remark, note that these equations have been derived heuristically in Ref.~\cite{pena2006coherent} by supposing that the exciting field is the average field transmitted through the monolayer. Instead, we show here that this result stems from a well-handled approximation that naturally appears in the multiple-scattering expansion.

\subsection{Generalization to particle monolayers on layered substrates}

We now move to the last step of the derivation, which is to consider the impact of a layered environment on the specular reflection and transmission of a particle monolayer. A first possible approach would be to start from Maxwell's equations and solve the electromagnetic scattering problem using the layered medium as a background. The dyadic Green function should then be the one of the layered geometry~\cite{paulus2000accurate}. While this could be done for very small particles behaving as electric dipoles~\cite{sasihithlu2016surface}, in which case scattering by an individual particle is not described by a non-local T-operator but by a local polarizability, it turns out to be much more challenging for large particles.

A more practical approach is to consider that the interaction with the layered geometry is mediated by the average scattered field from the particle monolayer \textit{in the uniform background}~\cite{garcia2012multiple}. In other words, the particle monolayer is treated as planar interface with reflection and transmission coefficients given by $r_\text{coh}$ and $t_\text{coh}$, and the reflection and transmission coefficients of the entire structure are determined by classical formulas of electromagnetic waves in layered media~\cite{yariv2005optical}. In addition to the approximations made at the level of the particle monolayer only, the mutual interaction between particles via the interfaces of the environment is therefore discarded in this approach, which may be problematic in certain situations, for instance, when incident planewaves couple efficiently to guided (photonic or plasmonic) modes in the layered medium.

In the framework of the latter approach and using classical recursive relations for wave propagation in thin films~\cite{yariv2005optical}, the reflection coefficient $r_\text{st}$ of a particle monolayer above a layered substrate (made of isotropic materials, to disregard polarization conversion\footnote{Multilayered substrates with anisotropic materials could be considered as well, in which case the reflection and transmission coefficients should be written as $2 \times 2$ matrices for TE and TM (or s and p) polarizations. Several publicly available codes can be used for the purpose~\cite{hugonin2021reticolo, bay2022pyllama}.}) is simply given by
\be
  r_\text{st} = r_\text{coh} + \frac{r_\text{sub} t_\text{coh}^2 \exp[2i k_0 n_\text{b} a \cos(\theta_\text{i})]}{1-r_\text{coh} r_\text{sub} \exp[2i k_0 n_\text{b} a \cos(\theta_\text{i})]},\label{eq:refl_stack}
\ee
where $r_\text{coh}$ and $t_\text{coh}$ are the reflection and transmission coefficients of the particle monolayer -- determined with any preferred model (ISA, EFA, QCA, ...), $r_\text{sub}$ is the reflection coefficient of the substrate alone, $n_\text{b} = \sqrt{\epsb}$ is the refractive index of the medium in which the particles are embedded, and $a$ is the height separating the monolayer and the first substrate interface ($a$ equals the particle radius for spherical particles). Importantly, $r_\text{sub}$ already takes into account all multiple reflections in-between the interfaces if several layers are considered. This quantity can be calculated using, for instance, the transfer matrix~\cite{mackay2020transfer} or the scattering matrix method~\cite{ko1988scattering}.

In the next section, we will test the ISA and EFA models for the specular reflection of particle monolayers on layered substrates in several experimentally-relevant situations.


\section{Numerical validation of theoretical predictions}
\label{sec:numerical-aspects}


To verify the validity of the ISA and EFA models for the problem of interest, we use an in-house multiple-scattering code developed by Jean-Paul Hugonin at Laboratoire Charles Fabry (Institut d'Optique Graduate School, CNRS, Universit\'e Paris Saclay). The numerical method belongs to the broad family of T-matrix methods~\cite{mishchenko1996t}, that is probably the most adapted to solve wave scattering problems by discrete media. T-matrix methods benefit from the high degree of analyticity of the functions describing the wave propagation between particles, typically dyadic Green functions or vector spherical wave functions, depending on the problem of interest and the specific implementation. With the recent emergence of disordered media in photonics~\cite{vynck2021light}, the modelling of multiple scattering by particles in layered media has gained considerable attention, leading to the development of new methods~\cite{bertrand2020global} and powerful publicly available tools~\cite{egel2021smuthi}.

One of the major difficulties in numerical studies of multiple scattering in disordered media comes from the fact that the phenomena of interest may take place on mesoscopic scales, much larger than the wavelength, implying that care should be taken to avoid finite-size effects. This issue is often circumvented partially by simulating larger systems illuminated by (smaller) collimated beams, but this strategy generally looses reliability for studies at large incident angles. 

\begin{figure}[h!]
	\centering
	\includegraphics[scale=1]{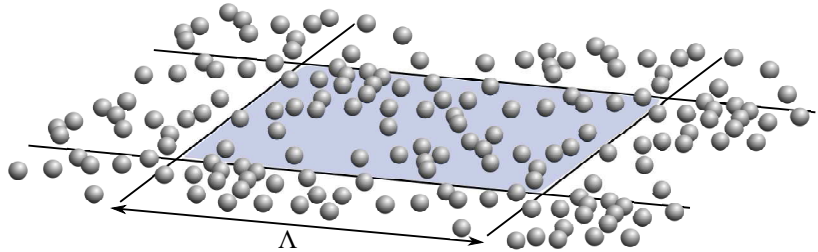}
	\caption{Sketch of the supercell approach used to compute the specular reflectance from particle monolayers on layered substrates. Non-overlapping particles are distributed, using a Random Sequential Addition algorithm, at random positions in a square of side $\Lambda$ with periodic boundary conditions. The unwanted effects of this artificial periodicity on the scattering properties are expected to vanish as $\Lambda$ increases.}
	\label{fig:supercell_sketch}
\end{figure}

In addition to being capable of modelling large particles incorporated in an arbitrary layered medium, made of isotropic or anisotropic materials, the in-house code used here implements the supercell method, that is an artificial periodization of the electromagnetic problem, wherein each supercell can contain a large number of particles [Fig.~\ref{fig:supercell_sketch}] to cope with finite-size effects. Technical details on the implementation can be found in Ref.~\cite{langlais2014cooperative}. The numerical code has been used successfully in several recent studies~\cite{jouanin2016designer, blanchard2020multipolar, bertrand2020global, vynck2022visual}.

To test the validity of the ISA and EFA models, we propose here to compute the specular reflectance spectra, $R_\text{st} = |r_\text{st}|^2$ of several systems under planewave illumination using the supercell approach. The artificial periodicity translates the scattering problem as a diffraction problem where the radiated power is distributed in several diffraction orders, with the 0-th order corresponding to the specular component and all the others to the diffuse component. One expects that, upon increasing the size of the supercell at constant particle density, the power radiated in the 0-th order converges towards a stable value.

\begin{figure}[h!]
	\centering
	\includegraphics[scale=1]{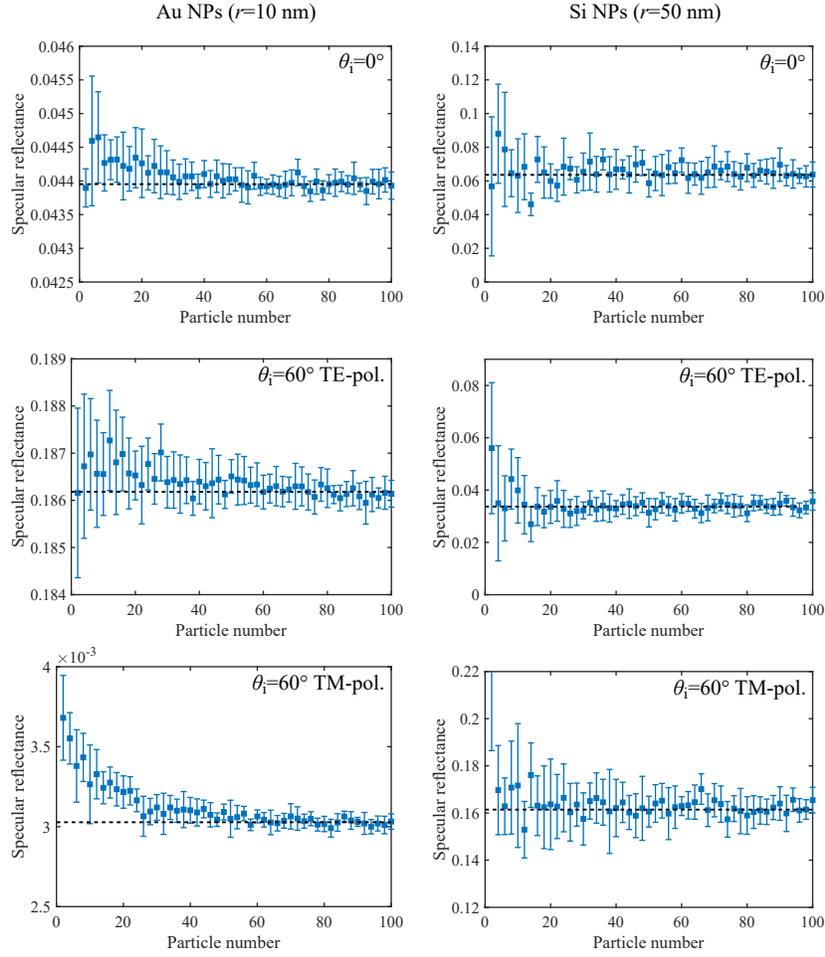}
	\caption{Convergence of the supercell method for the specular reflectance with increasing particle number. The convergence study is done in the same conditions as for the spectra calculations (incident angles and polarizations, number of disorder configurations). One expects to converge to a stabilized value as the particle number (or equivalently at fixed density, the system size) increases. The black dashed line serves as a guide to the eye. (Left) For a monolayer of spherical Au NPs on a SiO$_2$ substrate at a surface coverage $f=0.10$ and a wavelength $\lambda=500$ nm. (Right) For a monolayer of spherical Si NPs on a SiO$_2$ substrate at a surface coverage $f=0.10$ and at a wavelength $\lambda=470$ nm.}
	\label{fig:convergence}
\end{figure}

Let us then start our numerical study by testing the convergence of the method. We consider two systems for this convergence test, monolayers of either 10-nm-radius spherical gold (Au) particles or 50-nm-radius spherical silicon (Si) particles, deposited at a surface coverage of 10\% ($f=0.10$) on a semi-infinite silica (SiO$_2$) substrate. The particle monolayers are illuminated by a planewave at $\lambda=500$ and $470$ nm, respectively, and at either $\theta_\text{i}=0^\circ$ or $\theta_\text{i}=60^\circ$ in either TE or TM polarization. The simulations are repeated on 10 independent disorder configurations for systems up to $N=100$ particles. The results given in Fig.~\ref{fig:convergence} show some stabilization around a fixed value (the black dashed lines serve as guides to the eye), although some oscillations and fluctuations remain. Predictions appear reasonably accurate for $N\leq 50$.

\begin{figure}[h!]
	\centering
	\includegraphics[scale=1]{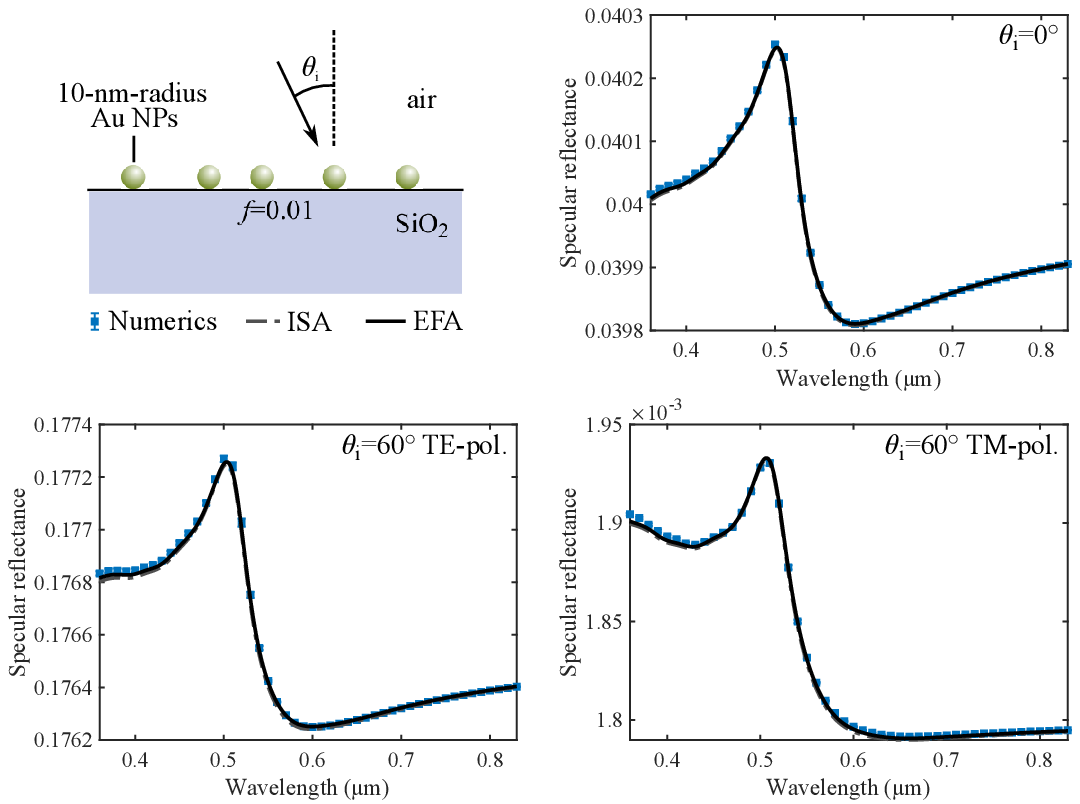}
	\caption{Specular reflectance spectra of a monolayer of 10-nm-radius gold (Au) nanoparticles (NPs) deposited at a surface coverage (or filling fraction) $f=0.01$ on top of a SiO$_2$ substrate. The three panels correspond to three different incident angles and polarizations: $0^\circ$ (polarization independent), $60^\circ$ in TE-polarization and in TM-polarization. The numerical predictions obtained by full-wave multiple-scattering computations (square markers, results averaged over 10 disorder configurations, error bar = standard deviation) are compared to the predictions from the ISA model (dotted-dashed line) and the ESA model (solid line).}
	\label{fig:spectrum_AuNPs_on_glass_low_ff}
\end{figure}

\begin{figure}[h!]
	\centering
	\includegraphics[scale=1]{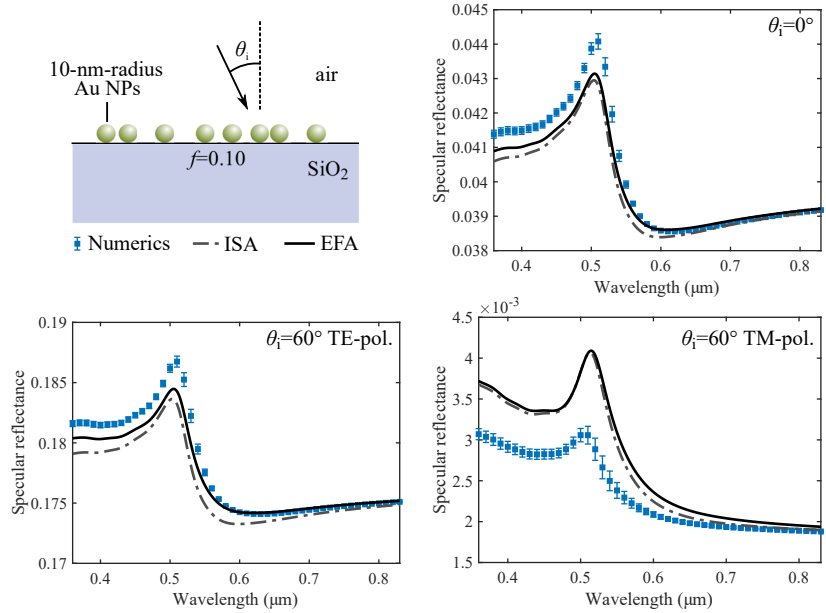}
	\caption{Same as Fig.~\ref{fig:spectrum_AuNPs_on_glass_low_ff} for a higher surface coverage $f=0.10$.}
	\label{fig:spectrum_AuNPs_on_glass_high_ff}
\end{figure}

\begin{figure}[h!]
	\centering
	\includegraphics[scale=1]{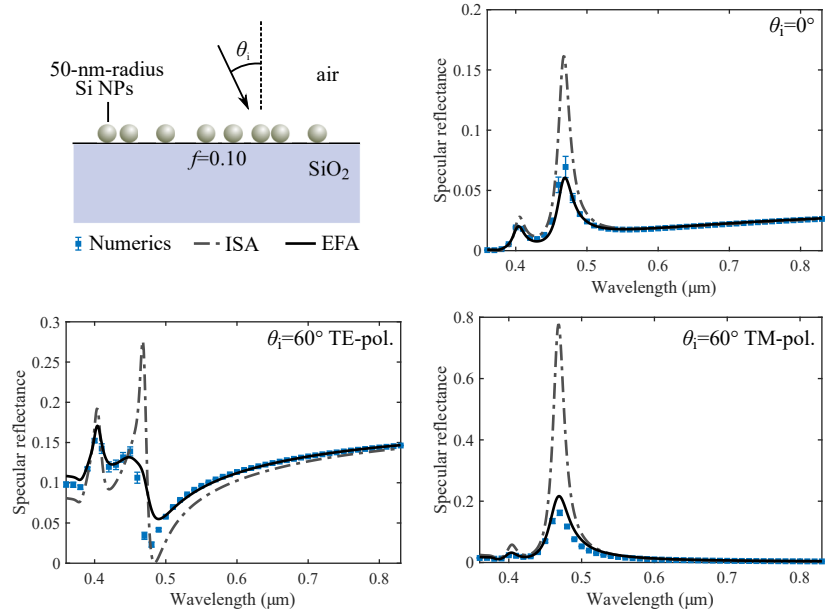}
	\caption{Same as Fig.~\ref{fig:spectrum_AuNPs_on_glass_high_ff} for 50-nm-radius silicon (Si) particles. The particles exhibit strong Mie resonances in the blue part of the spectrum.}
	\label{fig:spectrum_SiNPs_on_glass_high_ff}
\end{figure}

\begin{figure}[h!]
	\centering
	\includegraphics[scale=1]{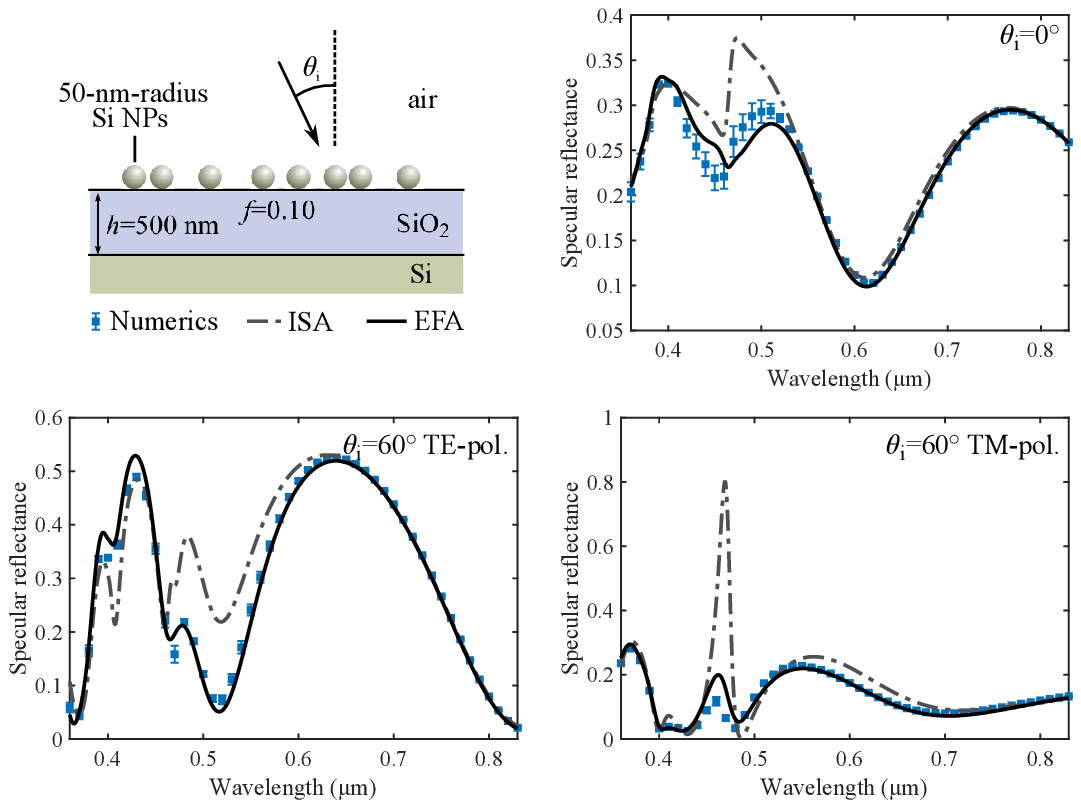}
	\caption{Same as Fig.~\ref{fig:spectrum_SiNPs_on_glass_high_ff} for a layered substrate composed of a 500-nm-thick SiO$_2$ layer on top of a semi-infinite Si substrate.}
	\label{fig:spectrum_SiNPs_on_SiO2_Si_high_ff}
\end{figure}

We thus proceed fixing $N=50$ for all systems and compute the specular reflectance spectra in the visible range for different systems, composed of spherical Au or Si particles at different surface coverages on different substrates. The results are given in Figs.\ref{fig:spectrum_AuNPs_on_glass_low_ff}-\ref{fig:spectrum_SiNPs_on_SiO2_Si_high_ff}. The numerical predictions are compared with those from the ISA model (Eqs.~\eqref{eq:refl_coeff_ISA}, \eqref{eq:trans_coeff_ISA} and \eqref{eq:refl_stack}) and the EFA model (Eqs.~\eqref{eq:refl_coeff_ISA}, \eqref{eq:trans_coeff_ISA} and \eqref{eq:refl_stack}).

For the gold particles at low surface coverage ($f=0.01$) [Fig.~\ref{fig:spectrum_AuNPs_on_glass_low_ff}], all model curves superimpose with the numerical data, showing clearly that the assumption of independent scattering is fully justified here. The peak observed in the spectrum corresponds to the plasmon resonance of individual particles.

The situation is different for the same system at higher surface coverage ($f=0.10$) [Fig.~\ref{fig:spectrum_AuNPs_on_glass_high_ff}], where the agreement between the models and the numerics is only moderately satisfactory. Remarkably, the EFA does not yield significantly better predictions compared to the ISA, suggesting that the mutual interaction between particles is more intricate than assumed. A reason may be the strong near-field interaction, accompanied by the formation of ``hot spots'', which are expected in dense heterogeneous metallic nanostructures.

The spectra of high-index Si particles at the same surface coverage ($f=0.10$) [Fig.~\ref{fig:spectrum_SiNPs_on_glass_high_ff}] exhibit sharp spectral resonances due to the Mie resonances of the individual particles. The strength of these resonances are yet strongly overestimated by the ISA model, especially at grazing angles, where values up to about 4 times the numerical value are reached. This is likely being due to the unphysical divergence of the coefficients at large angles in the ISA. By comparison, the EFA model performs very well: despite some inaccurate predictions near resonance wavelengths, the spectral features are overall very well reproduced.

Even more impressive are the results for the same particle monolayer on a layered substrate composed of a 500-nm-thick SiO$_2$ intermediate layer of top of a semi-infinite Si substrate. The multiple reflections between the monolayer and the various interfaces lead to strong spectral variations, which are well captured by the EFA, contrary to the ISA.

All in all, our full-wave computations show that, whereas the use of the ISA should be restricted to very dilute systems, the EFA model can be quantitatively accurate for particle monolayers with surface coverages of about 10\% even at large angles of incidence. The specular reflection and transmission coefficients from particle monolayers have been derived here, step by step, from Maxwell's equations. We hope that this derivation and the numerical tests will be useful to the scientific community.

\section{Acknowledgments}
\label{sec:acknowledgments}

We are grateful to Jean-Paul Hugonin (Laboratoire Charles Fabry, Palaiseau, France) for providing the multiple-scattering code used in this work. This work has received financial support from the PSA group and from the french National Agency for Research (ANR) under the projects ``NanoMiX'' (ANR-16-CE30-0008) and ``NANO-APPEARANCE'' (ANR-19-CE09-0014).



\end{document}